%% file: main.tex
\documentclass[10pt,conference]{IEEEtran}
\IEEEoverridecommandlockouts

\usepackage{cite}
\usepackage{amsmath,amssymb,amsfonts}
\usepackage{textcomp}
\usepackage{xcolor}
\usepackage[most]{tcolorbox}
\usepackage{balance}
\usepackage{enumitem}
\usepackage{hyperref}
\usepackage{listings} 
\usepackage{xcolor}
\usepackage{microtype} 
\usepackage{xspace}
\usepackage{algorithm}
\usepackage{algpseudocode}
\usepackage{stmaryrd}
\usepackage{soul}
\usepackage{color, colortbl}
\usepackage{amsfonts} 
\usepackage{booktabs} 
\usepackage{graphicx, subfigure}
\usepackage{caption}
\usepackage{epigraph} 
\usepackage{framed}
\usepackage{here}
\definecolor{coolblack}{rgb}{0.0, 0.18, 0.39}
\usepackage{multirow}
\usepackage{xurl}
\usepackage{comment}
\usepackage{soul}
\usepackage{here}

\newcommand{\RQone}{RQ1 -- \emph{How promptly do packages integrating Log4j-Core address CVEs?}\xspace}
\newcommand{\RQoneline}{\ul{RQ1 -- \emph{How promptly do packages integrating Log4j-Core address CVEs?}}\xspace}
\newcommand{\RQtwo}{RQ2 -- \emph{What factors influence the response time to critical CVEs in packages using Log4j-Core?}\xspace}

\newcommand{\RQtwooneline}{\ul{RQ2.1 -- \emph{To what extent is release frequency associated with the response time to critical CVEs?}}\xspace}
\newcommand{\RQtwotwoline}{\ul{RQ2.2 -- \emph{To what extent do response times to critical CVEs vary across major, minor, and patch versions?}}\xspace}

\definecolor{awesome}{rgb}{0.0, 0.2, 0.6}
\input{comments}

\def\BibTeX{{\rm B\kern-.05em{\sc i\kern-.025em b}\kern-.08em
    T\kern-.1667em\lower.7ex\hbox{E}\kern-.125emX}}
    
\begin{document}

\title{Mining for Lags in Updating Critical Security Threats: A Case Study of Log4j Library}

\author{\IEEEauthorblockN{Hidetake Tanaka$^1$, Kazuma Yamasaki$^1$, Momoka Hirose$^1$, Takashi Nakano$^1$, \\ Youmei Fan$^1$, Kazumasa Shimari$^1$, Raula Gaikovina Kula$^2$, Kenichi Matsumoto$^1$}
	\IEEEauthorblockA{
		$^1$\textit{Graduate School of Science and Technology, Nara Institute of Science and Technology}\\
        $^2$\textit{Graduate School of Information Science and Technology, Osaka University}\\
		\{tanaka.hidetake.te0, yamasaki.kazuma.yj9, hirose.momoka.hm4\}@naist.ac.jp, \\
        \{nakano.takashi.nr1, fan.youmei.fs2, k.shimari, matumoto\}@is.naist.jp,\\ raula-k@ist.osaka-u.ac.jp}
}

\maketitle
\begin{abstract}
The Log4j-Core vulnerability, known as Log4Shell, exposed significant challenges to dependency management in software ecosystems. When a critical vulnerability is disclosed, it is imperative that dependent packages quickly adopt patched versions to mitigate risks. However, delays in applying these updates can leave client systems exposed to exploitation.
Previous research has primarily focused on NPM, but there is a need for similar analysis in other ecosystems, such as Maven. Leveraging the 2025 mining challenge dataset of Java dependencies, we identify factors influencing update lags and categorize them based on version classification (major, minor, patch release cycles).
Results show that lags exist, but projects with higher release cycle rates tend to address severe security issues more swiftly. In addition, over half of vulnerability fixes are implemented through patch updates, highlighting the critical role of incremental changes in maintaining software security. Our findings confirm that these lags also appear in the Maven ecosystem, even when migrating away from severe threats.
\end{abstract}


\section{Introduction}
Open-source software forms the backbone of modern software development, enabling rapid innovation and cost-effective solutions. Developers heavily rely on open-source libraries and frameworks to integrate pre-built functionalities, significantly reducing development time and resources. 
Despite these advantages, this dependence presents critical challenges, particularly in managing dependencies and addressing security vulnerabilities. While dependencies are essential for productivity and code reuse, they can introduce significant security risks when widely used libraries contain critical vulnerabilities~\cite{Liu2020ICSE, Kula2018EMSE, Lauinger2017NDSS}. When such vulnerabilities are disclosed, dependent packages must swiftly adopt patched versions to mitigate potential threats. However, delays—referred to as update lags—in applying these patches can leave systems exposed to exploitation, increasing the overall vulnerability of software ecosystems~\cite{Decan2018ICSME}.

A recent and widely discussed example is the Log4j-Core library, which gained global attention due to the Log4Shell vulnerability~\cite{Hiesgen2024The}. This critical vulnerability demonstrated the far-reaching impact that a single flaw in a widely used library can have, jeopardizing systems across various industries, including enterprise software and critical infrastructure. To address the vulnerability, the patched version Log4j-Core 2.17.0 was released on December 17, 2021. However, the speed with which the dependent packages adopted this update varied significantly, raising important questions about the responsiveness of dependency updates and the factors that influence these behaviors.

Several studies have focused on the exploitation and mitigation of the Log4Shell vulnerability. Feng and Lubis~\cite{Feng2022Defense} proposed a defense-in-depth security strategy to analyze and mitigate Log4Shell, emphasizing layered approaches to protect systems against similar threats. Kaushik et al.~\cite{Kaushik2022An} examined specific exploitation methods for Log4Shell and proposed mitigation techniques, providing a practical foundation for understanding the risks and countermeasures associated with such vulnerabilities. Juvonen et al.~\cite{Juvone2022On} investigated the impact of the Log4Shell vulnerability on critical communication systems, including aeronautical and maritime domains, highlighting the far-reaching consequences of delayed updates in systems with stringent reliability requirements. Hiesgen et al.~\cite{Hiesgen2022The} measured response times to the Log4Shell incident, identifying patterns in how quickly vulnerabilities were exploited after their disclosure.
From a broader perspective, several works have addressed dependency management and version updates in open-source ecosystems. Zhang et al. \cite{Zhang2023Mitigating} analyzed the persistence of vulnerabilities in Maven dependencies and proposed strategies to enhance security by improving update practices. Bao et al.~\cite{Bao2022V-SZZ} introduced V-SZZ, a method to automatically identify version ranges affected by CVE vulnerabilities, enabling efficient tracking of security issues across software versions. Sopariwala et al.~\cite{Sopariwala2022Log4jPot} developed Log4jPot, a detection system for identifying Log4Shell vulnerabilities, emphasizing the importance of detection as a precursor to timely updates.

\begin{figure*}[t]
   \centering
	\includegraphics[width=1.4 \columnwidth]{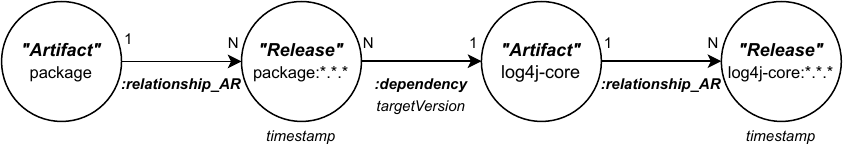}
    \caption{The structure of the data extracted from Neo4j.}
	\label{fig:neo4j_struct}
\end{figure*}

While these studies provide valuable insights into the challenges and solutions surrounding dependency management and vulnerability mitigation, few have specifically addressed the update lag in dependent packages or analyzed the factors influencing update responsiveness. This lag is critical, as timely adoption of security patches in dependent packages is essential to protect software ecosystems from exploitation.

To address this lag, our study investigates the responsiveness of packages dependent on Log4j-Core to the release of its critical patch.
One of the key goals is to understand the relationship between the release cycles of a client system (i.e., major, minor and patch) and the how quickly the maintainers migrated away from the vulnerability.
Specifically, we focus on the following research questions:

\textbf{\RQone} 
This research question aims to clarify how promptly packages address critical CVEs. As a case study, we empirically examine the time it took for packages integrating Log4j-Core to update their versions to 2.17.0 or later.

\textbf{\RQtwo} To answer this research question, we split two sub-research questions.

\RQtwooneline This research question examines the responsiveness of dependent packages in adopting critical updates. We analyze how often packages dependent on Log4j-Core integrate the patched version (Log4j-Core 2.17.0). This analysis identifies patterns in release frequency and highlights characteristics of packages that tend to respond promptly to security vulnerabilities.

\RQtwotwoline By categorizing updates of dependent packages into major, minor, and patch changes, this question explores how the type of update influences the timeliness of dependency updates. We investigate whether smaller updates (e.g., patches) are adopted more promptly than larger updates (e.g., major or minor), and we analyze the patterns of update lag across these classifications. Since patch updates are generally smaller and simpler than major ones, the associated time with such updates also tends to be smaller. This analysis identifies that trend.

Through these research questions, this study aims to uncover actionable insights into the factors influencing dependency update behaviors. The findings contribute to a deeper understanding of dependency management challenges and offer practical recommendations for improving security practices in software ecosystems.
For reproducibility of experiments, the replication package is available in the GitHub repository.\footnote{\url{https://github.com/NAIST-SE/MSR2025-log4shell-depend-analyze}}

\section{Study Design}
In this section, we present the data collection.
We collect dependency data on the Log4j-Core releases and the dependency relationship of packages depending on Log4j-Core from the Neo4j dataset generated by Goblin Miner~\cite{Jaime2025GOBLIN}.

\subsection{Data Preparation and Extraction}
The structure of the data extracted from Neo4j is shown in the Figure~\ref{fig:neo4j_struct}.
This figure illustrates the relationships between artifacts and their dependencies, as represented in the Neo4j database.
In data structures, “1 → N” indicates a one-to-many relationship where a single element (1) is associated with multiple elements (N), while “N → 1” represents a many-to-one relationship where multiple elements (N) are associated with a single element (1).
Each \textit{Artifact} node represents a software component (e.g., Log4j-Core), while \textit{Release} nodes represent specific versions of these artifacts. The ``targetVersion'' attribute of \textit{dependency} relationships indicate which version of Log4j-Core is used by a given package. This data structure is critical for analyzing how quickly dependent packages adopt patched versions of Log4j-Core. 
We extract the releases of packages depending on Log4j-Core from the dataset.
Specifically, we identify \textit{Release} nodes connected to the \textit{Artifact} node with the ``id'' property ``org.apache.logging.log4j:log4j-core'' through relationships labeled as ``dependency''.

For data extraction, we limit our selection to versions that strictly adhere to pure semantic versioning.
Specifically, we include versions represented in the format ``Major.Minor.Patch'', where numeric components are separated by dots (e.g., 2.17.1).
Conversely, versions such as ``*.*.*-beta” or “*.*.*-alpha'' (where * represents any numeric component), which contain additional labels (e.g., pre-release identifiers or metadata), are excluded as they are not considered production releases\footnote{https://semver.org/}.
Additionally, we exclude \textit{Release} nodes with \textit{dependency} relationships whose ``targetVersion'' property is specified as a range or is otherwise not a specific version. This is because it is not possible to determine which exact version they depend on.
We collect data on 10,650 artifacts and 402,232 releases.
For each release, we collect information on the release version and its timestamp, as well as the version of Log4j-Core it depends on and its timestamp. This data is then grouped by artifact for further processing.

To identify which client systems did migrate away from the vulnerable dependency, we identify and extract packages from the collected data that had been updated to version 2.17.0 or later.
Specifically, each artifact arranges its releases in chronological order. When a release with a version below 2.17.0 appears followed by a release with a version 2.17.0 or higher, the artifact is considered updated to 2.17.0 or later.
For each extracted data point, the version and release timestamp of the releases before and after updating the dependency of Log4j-Core across 2.17.0, as well as the version of Log4j-Core after the update and its release timestamp, are extracted as data.
We collect data on 2,210 updates.

\begin{figure}[tb]
   \centering
	\includegraphics[width=0.8\columnwidth]{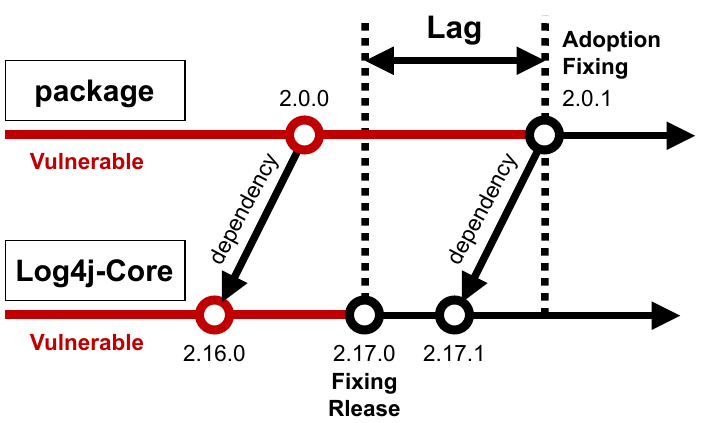}
    \caption{Overview of the Lag between the fixing of the Log4j-Core vulnerability and the adoption of the fix to the package.}
	\label{fig:lag}
\end{figure}

\subsection{Empirical Study Design}

\RQoneline

As a measure of how quickly do packages respond, we define ``Lag'' as the number of days elapsed from the release of Log4j-Core version 2.17.0 to the point when a package updates its dependency to 2.17.0 or later.
The period between when the vulnerability is fixed in the dependency package and when the fix is applied to the dependent package is referred to as the ``Lag''. For analysis, the Lag is calculated for all artifacts, aggregated, and represented as a histogram.

Figure~\ref{fig:lag} illustrates an overview of the Lag. The 2.0.0 version of a package depends on Log4j-Core 2.16.0. Since Log4j-Core:2.16.0 still has the Log4Shell vulnerability, the package:2.0.0 is still vulnerable. Later, Log4j-Core fixed the vulnerability, but since the package:2.0.0 did not update its dependencies, the vulnerability remained. Subsequently, the package updated its dependency to Log4j-Core:2.17.1 or later, resolving the vulnerability in the package.

\RQtwooneline
To measure release frequency, we use a metric that indicates how many days, on average, each artifact takes to produce a new release.
Specifically, we employ the method described in paper~\cite{Ruiz2023A}.
We calculate this metric by dividing the time difference between the timestamps of its first and last releases by the total number of releases (minus 1) for an artifact when arranged in chronological order.
For each artifact, a scatter plot is constructed with Lag on the horizontal axis and release frequency on the vertical axis.

\RQtwotwoline
We categorize updates based on which version component—Major, Minor, or Patch—was updated, and measure the Lag for each category.
The determination of which version component was updated is based on the changes in the version values of Log4j-Core dependencies before and after the update across 2.17.0. If multiple version components (e.g., Major, Minor, Patch) change, the update is categorized according to the highest-priority component, with the order of precedence being Major $>$ Minor $>$ Patch.
The Lag is aggregated for each version component category (Major, Minor, Patch) and presented as a box plot.

\begin{figure}[t]
   \centering
	\includegraphics[width=\columnwidth]{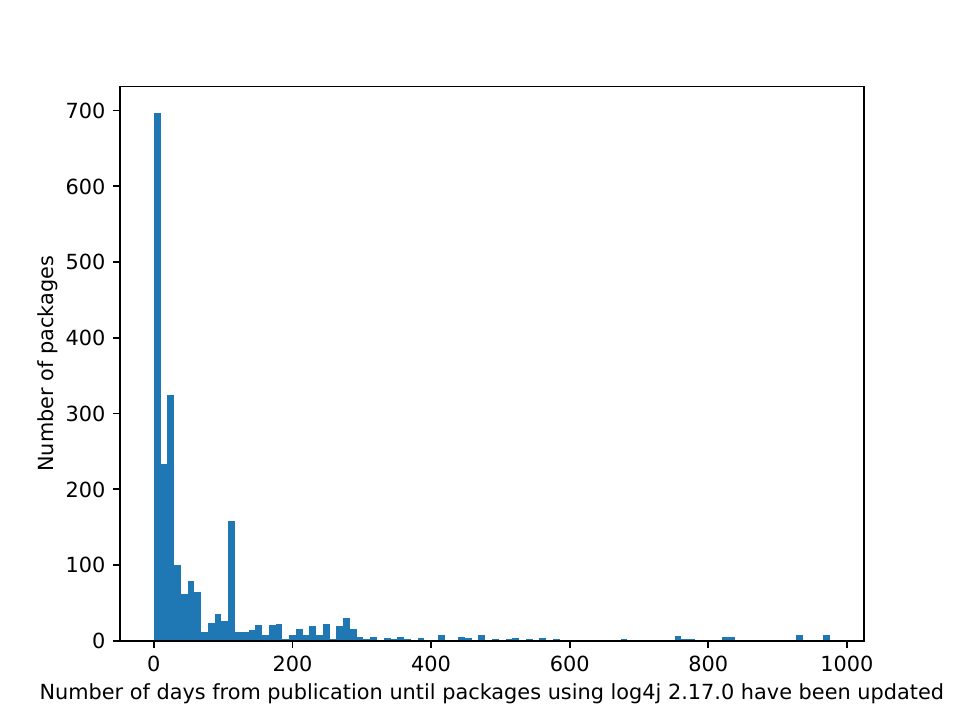}
    \caption{Relationship between the number of days to respond to the CVE and the number of packages.}
	\label{fig:rq1}
\end{figure}

\section{Results}
\subsection{\textbf{\RQone}}
Figure~\ref{fig:rq1} shows the results of RQ1, showing the distribution of the number of days it took for packages to update their dependency on Log4j-Core to the patched version Log4j-Core 2.17.0. For example, in the figure, we show that the majority of packages updated within the first three months following the release of the patched version, with a smaller but notable proportion taking up to a year to update. 

We highlight two findings.
The first finding from the analysis is that the majority of packages demonstrated a relatively quick response to the disclosed CVE. Specifically, 72.67\% of packages updated within the first three months, indicating a significant awareness and effort to address the vulnerability promptly.
The second finding from the analysis is that while 95.07\% of packages eventually updated within one year, the remaining 4.93\% exhibited substantial delays or failed to respond altogether, raising concerns about the persistence of vulnerable versions in the ecosystem.

\begin{tcolorbox}[colback=gray!5,colframe=awesome,title= RQ1 Summary]
There is a persistence of vulnerable versions in the ecosystem.
Results show that 72.67\% of packages updated to the patched version of Log4j-Core within the first three months, demonstrating a prompt response to the vulnerability. By one year, 95.07\% of packages had updated, while 4.93\% experienced significant delays or failed to update. 

\end{tcolorbox}

\noindent\RQtwooneline

\begin{figure}[t]
   \centering
	\includegraphics[width=\columnwidth]{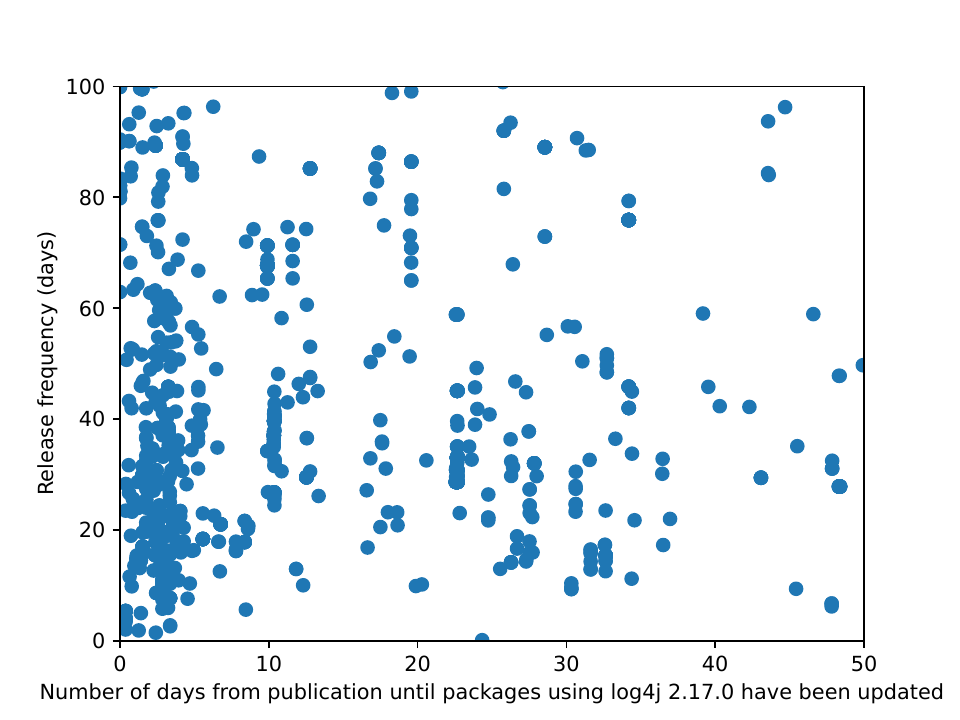}
    \caption{Relationship between the number of days to respond to the CVE and the frequency of releases.}
	\label{fig:rq2_1}
\end{figure}

Figure \ref{fig:rq2_1} visualizes the relationship between release frequency and the time taken to update in response to the critical CVE. The analysis revealed a positive correlation (correlation coefficient of 0.43), indicating that packages with higher release frequencies tended to respond more quickly to the vulnerability.
This finding suggests that frequent updates may be indicative of an active maintenance process, which contributes to faster responses to critical security issues. In other words, organizations that frequently release new versions of their software tend to address security vulnerabilities more promptly than those that do not.

By examining the relationship between release frequency and response time to CVEs, we gain insights into the dynamics of software maintenance and updates. The observed positive correlation highlights the importance of regular releases in ensuring timely responses to critical security issues. This finding has implications for software development practices, highlighting the need for organizations to prioritize frequent releases and active maintenance to stay ahead of emerging security threats.


\subsection{\textbf{\RQtwo}}

\begin{figure}[t]
   \centering
	\includegraphics[width=\columnwidth]{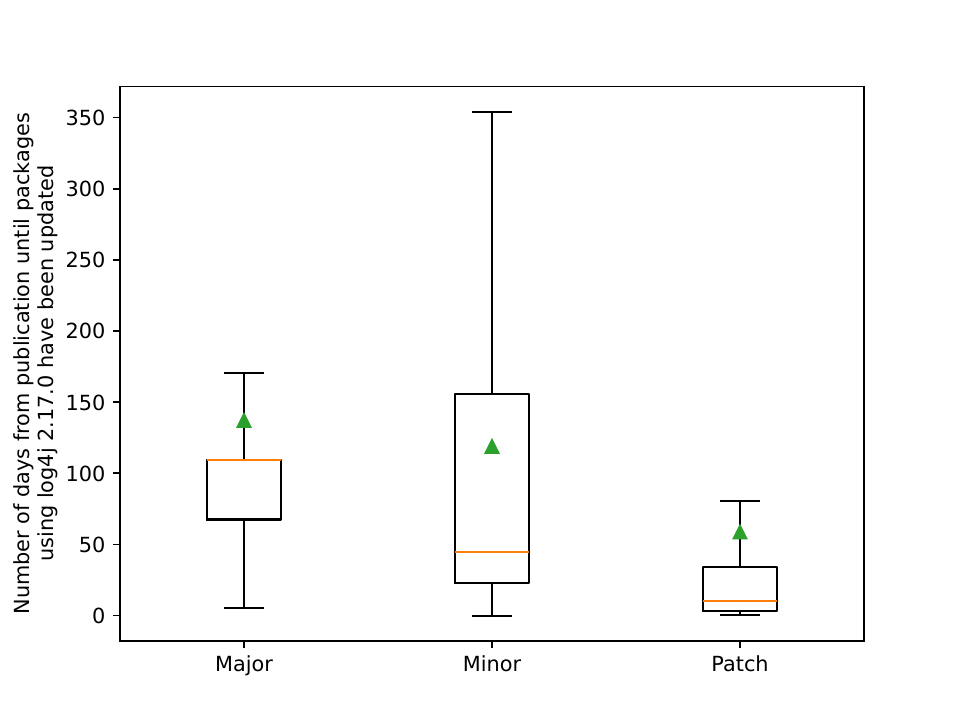}
    \caption{Relationship between the semantic versioning part and the number of days to respond to the CVE.}
	\label{fig:rq2_2}
\end{figure}

\noindent\RQtwotwoline

Figure~\ref{fig:rq2_2} shows the differences in response times based on the type of semantic versioning update (major, minor, or patch). Packages that updated using patch versions responded the fastest, with a median response time of 10 days. In contrast, updates involving major version changes showed the longest response times, with a median of 109 days. This result indicates that minor and patch updates are more likely to be applied promptly in response to critical CVEs, whereas major updates may involve additional complexity or testing requirements.

\begin{tcolorbox}[colback=gray!5,colframe=awesome,title= RQ2 Summary]
Packages that update frequently also respond faster to critical CVEs.
\begin{itemize}[leftmargin=1em]
\item Answering \textbf{RQ2.1}, we find there is a moderate positive correlation (correlation coefficient of 0.43) between time spent updating and release frequency.
\item Answering \textbf{RQ2.2}, we find that the mean and median number of days required to respond to a critical CVE relate to the release cycle.
\end{itemize}
\end{tcolorbox}

\section{Discussion and Future Work}

As part of this mining challenge, we found that even critical vulnerabilities like the Log4Shell vulnerability took around three months to be addressed, which is a concerning delay.

Using the dataset from the mining challenge, we gained insights into the update latency of vulnerabilities in the Maven software ecosystem. We observed that projects with higher update frequencies exhibited lower update latencies.

As part of our future work, we plan to target projects with lower update latencies and investigate the reasons behind their ease of updating compared to those with high lags. This could involve analyzing factors such as team size, communication patterns, or project governance structures.
Additionally, we can expand on our dataset by incorporating additional meta-information, such as the type of project (e.g., library, framework, plugin, web application, etc ), number of contributors, stars, and other relevant metrics, to better understand which projects tend to exhibit low update latencies. By exploring these factors, we may uncover patterns or trends that can inform best practices for vulnerability management and software maintenance.

\section*{Acknowledgment}
This work has been supported by JSPS KAKENHI Nos. JP20H05706, JP23K28065, JP23K16862, JP24K14895, and JST BOOST Grant Number JPMJBS2423.

\bibliographystyle{IEEEtranS.bst}
\bibliography{reference}

\end{document}

%% file: comments.tex
\usepackage{datenumber}

\definecolor{chestnut}{rgb}{0.8, 0.36, 0.36}

\definecolor{chestnut}{rgb}{0.8, 0.36, 0.36}

\newcounter{dateone}\newcounter{datetwo}%
\newcommand{\daydifftoday}[3]{%
\setmydatenumber{dateone}{\the\year}{\the\month}{\the\day}%
\setmydatenumber{datetwo}{#1}{#2}{#3}%
\addtocounter{datetwo}{-\thedateone}%
\thedatetwo
}